\documentclass{moriond}

\def\Journal#1#2#3#4{{#1} {\bf #2}, #3 (#4)}

\def\PRL{\em Phys. Rev. Lett.}
\def\PRD{{\em Phys. Rev.} D}

\def\be{\begin{equation}}
\def\ee{\end{equation}}
\def\bea{\begin{eqnarray}}
\def\eea{\end{eqnarray}}

\newcommand{\Photo}{}

\begin{document}
\vspace*{4cm}
\title{Probing flavor effects in QCD showers with heavy-flavor jets}

\author{ Emma Yeats on behalf of the ALICE Collaboration}

\address{Department of Physics, University of California Berkeley,\\
Berkeley, California, 94709, USA}

\maketitle\abstracts{
Measurements of jet substructure provide precise tests of Quantum Chromodynamics (QCD) and offer a distinct way to study hadronization mechanisms, compared to measurements of hadrons alone. QCD predicts that jet radiation patterns depend on the mass and color charge of the initiating parton. Parton showers, in particular, are sensitive to the Casimir factors of quarks and gluons, as well as the parton mass due to the dead-cone effect. Three key charm-tagged jet measurements from the ALICE experiment are discussed.}

\section{Introduction to Heavy-Flavor Jets}

Heavy-flavor quarks are often produced in the initial hard scattering and can be traced to the final-state hadrons we observe in our detectors. Jets tagged with heavy-flavor quarks in proton--proton (pp) collisions exhibit two interesting shower properties that distinguish them from inclusive (gluon-dominated at LHC energies) or light-flavor jets. The first is called the dead-cone effect, a suppression of gluon radiation in a cone, $\theta \approx m_{\rm Q}/E_{\rm Q}$, around the direction of the radiating quark. This suppression decreases with quark energy ($E_{\rm Q}$), as observed by the ALICE experiment in 2023 \cite{deadcone}, and increases with quark mass ($m_{\rm Q}$). The second property of heavy-flavor jets is due to the difference in Casimir color factors between quarks and gluons, which leads to quark-initiated showers exhibiting a harder and more collinear fragmentation profile than gluon-initiated showers. 

Three measurements of jet substructure observables are highlighted to examine these shower properties of heavy-flavor jets: Energy-Energy Correlators (EECs), defined as the energy-weighted angular cross section between particle pairs; jet axes differences, the opening angle between various jet axis definitions; and $z_{\rm g}$, the shared momentum fraction of a splitting.

Charm-jets are tagged by the presence of a $\rm D^0$ reconstructed from the $\rm D^0 \rightarrow K^{-} \pi^+$ (and charge conjugate) decay chain, which has a branching ratio of $(3.947 \pm 0.030)\%$ \cite{PDG}. When the $\rm D^0$ decay is identified, the daughter tracks are replaced with the corresponding $\rm D^0$ candidate, which is assigned the four-momentum sum of the daughters. We reconstruct track-based jets with the anti-$k_{\rm T}$ algorithm \cite{anti-kT} ($R=0.4$).

\section{The $\rm D^0$-tagged Energy-Energy Correlator (EEC)}

The EEC is the energy-weighted two-particle correlation inside jets and is given by \cite{eec}:
\begin{equation}
\Sigma_{\rm EEC}(R_{\rm L}) = \frac{1}{N_{\rm jet}\Delta}\sum_{N_{\rm jet}}\int\sum_{i,j} \left( \frac{p_{{\rm T}, i} p_{{\rm T}, j}}{p_{\rm T, jet}^2} \right) \delta(R_{\rm L}' - R_{{\rm L}, ij}){\rm d}R_{\rm L}',
\end{equation}
where the summation runs over the total number of jets, $N_{\rm jet}$, and all final-state particle pairs ($i,~j$) inside each jet. The angular interval size is denoted by $\Delta$. The energy-weight term utilizes the transverse momentum of the pairs over $p_{\rm T, jet}^2$ to suppress the contribution from soft tracks, restoring IRC safety to this observable and allowing for comparison to precise theoretical predictions. Using the opening angle between each pair, $R_{{\rm L}} = \sqrt{\Delta \varphi_{ij}^2 + \Delta \eta_{ij}^2}$, the EEC is uniquely suited to separate perturbative from non-perturbative physics. Specifically, large $R_{\rm L}$ is correlated to early splittings (more perturbative), while small $R_{{\rm L}}$ is correlated to late splittings (more non-perturbative). The transition region, or ``peak", between them corresponds to the onset of hadronization and is sensitive to the mass of the quark initiating the jet \cite{kyle_lee_EEC}.

\begin{figure}
\centerline{\includegraphics[width=0.425\linewidth]{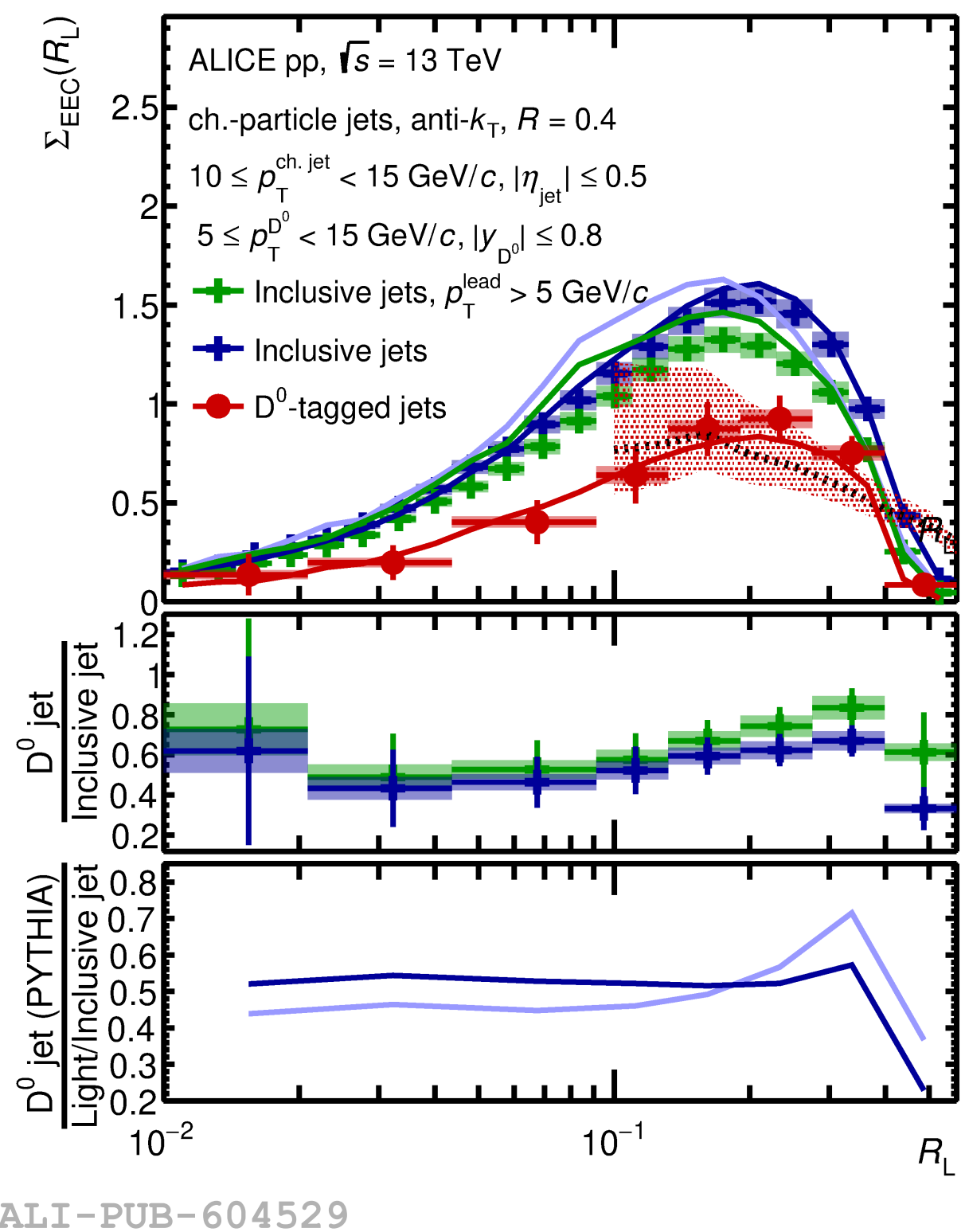}}
\caption[]{The $\rm D^0$-tagged jet EEC distribution (red) compared to inclusive jets, with (green) and without (blue) a leading track-$p_{\rm T}$ selection of $5~{\rm GeV}/c$, in the $10 < p_{\rm T}^{\rm ch~jet} < 15~{\rm GeV}/c$ interval. Comparisons to PYTHIA 8 simulations and pQCD predictions are also shown \cite{eec}.} 
\label{fig:eec}
\end{figure}

Recent results from this measurement are shown in Fig.~\ref{fig:eec}. A comparison between the inclusive-jet EEC, constructed from predominantly ($\sim76\%$) gluon-initiated jets \cite{eec}, to the light-quark jet prediction from PYTHIA 8, where the dead-cone is negligible, shows a peak shift to small $R_{\rm L}$ corresponding to the effect of Casimir factors. In the $\rm D^0$-tagged EEC, one sees a reduced total integral and thus a significant suppression across the whole range due to the dead-cone. The peaks of the $\rm D^0$-tagged and inclusive EECs are suprisingly similar, as the dead-cone effect shifts the $\rm D^0$-tagged EEC toward large $R_{\rm L}$ by suppressing small-angle splittings, while the larger Casimir color factor of gluons leads to the broadening of inclusive jets and also shifts the inclusive EEC to large $R_{\rm L}$. Calculations from perturbative QCD (pQCD) reproduce the general shape at large $R_{\rm L}$, but there is some tension near the peak. However, predictions from PYTHIA 8 describe the inclusive and $\rm D^0$-tagged measurements within uncertainties.

\section{The $\rm D^0$-tagged Jet Axes Differences}
\label{sec:jetaxes}
Jet Axes Differences are defined as the opening angle between jet axes \cite{jetaxes}, 
\begin{equation}
\Delta R_{\rm axis} = \sqrt{\Delta \varphi^2 + \Delta \eta^2}.
\end{equation}
We study three jet axes, described now in order of decreasing sensitivity to soft radiation. First, the Standard (STD) axis results from clustering the constituents of anti-$k_{\rm T}$ jets using $E$-scheme, which calculates the jet axis using the sum of the constituents' four-momenta. We then recluster the anti-$k_{\rm T}$ jets with the Cambridge-Aachen algorithm, which orders the splittings of the jet by angle. To create the Soft Drop (SD) axis, we apply the Soft Drop \cite{SD} condition,
\begin{equation}
\frac{min(p_{\rm T_1}, p_{\rm T_2})}{p_{\rm T_1}+p_{\rm T_2}} > z_{\rm cut} \left( \frac{\Delta R_{12}}{R} \right) ^\beta,
\label{eq:SD}
\end{equation}
where $\beta$ is set to 0 such that we only remove radiation softer than the value of $z_{\rm cut}$, a tuneable parameter. Eq.~\ref{eq:SD} is checked at each splitting following the harder prong, and the first splitting that satisfies it defines the jet axis. Lastly, the Winner-Takes-All (WTA) jet axis is calculated by recombining the angular-ordered jets with the Winner-Takes-All recombination scheme, which aligns the axis with the hardest subjet in the hardest branch at each clustering step \cite{wta}.

\begin{figure}
\centerline{\includegraphics[width=0.445\linewidth]{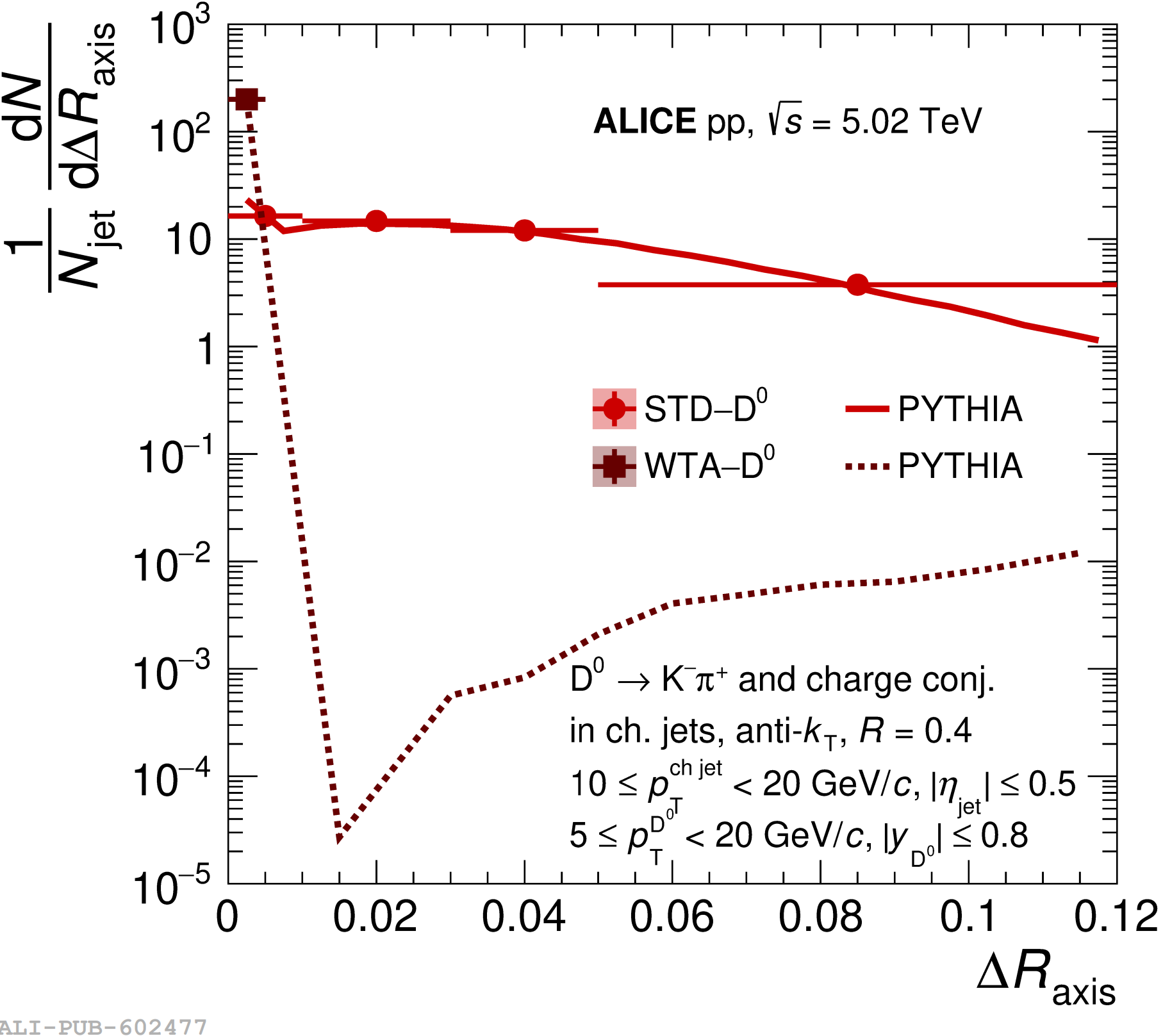} 
\includegraphics[width=0.44\linewidth]{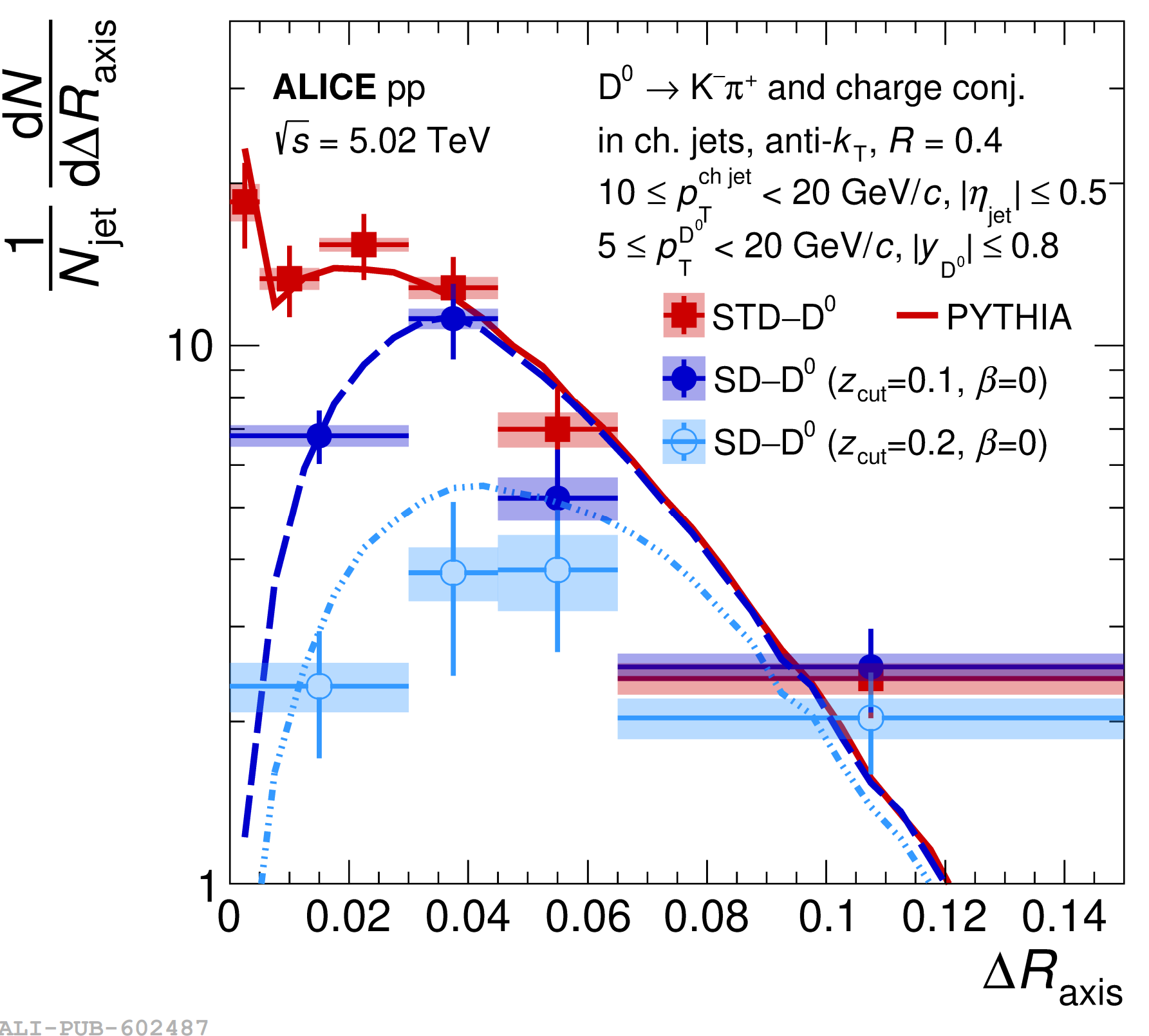}}
\caption[]{$\rm D^0$-tagged $\Delta R_{\rm STD-D^0}$ (red) compared to $\Delta R_{\rm WTA-D^0}$ (brown) in Fig. \ref{fig:jetaxes}(a), and to $\Delta R_{\rm SD-D^0}$ in Fig. \ref{fig:jetaxes}(b) with $z_{\rm cut}=0.1$ (dark blue) and $z_{\rm cut}=0.2$ (light blue) in the $10 < p_{\rm T}^{\rm ch~jet} < 20~{\rm GeV}/c$ interval with a $5 < p_{\rm T}^{\rm D^0} < 20~{\rm GeV}/c$ selection applied. The measurements are compared to predictions from PYTHIA 8 \cite{jetaxes}.} 
\label{fig:jetaxes}
\end{figure}

From the opening angles between these jet axes and the $\rm D^0$-meson direction, we can define several jet axes differences. In Fig.~\ref{fig:jetaxes}(a), $\Delta R_{\rm WTA-D^0}$ is shown and compared to $\Delta R_{\rm STD-D^0}$ as a reference. The extremely strong alignment between the Winner-Takes-All and $\rm D^0$ axes is due to the sharp peak near $\Delta R_{\rm axis} = 0$. We find that $(99\pm 1)\%$ of jets occupy this smallest $\Delta R_{\rm axis}$ interval when $10 < p_{\rm T}^{\rm ch~jet} < 20~{\rm GeV}/c$ and $p_{\rm T}^{\rm D^0} > 5~{\rm GeV}/c$. Because the Winner-Takes-All axis is strongly correlated to the leading particle in the jet, this alignment implies that the $\rm D^0$ meson is the leading particle (in the hardest prong) in all $\rm D^0$-tagged jets in the kinematic range. 

In Fig.~\ref{fig:jetaxes}(b), $\Delta R_{\rm STD-D^0}$ is compared to $\Delta R_{\rm SD-D^0}$ ($z_{\rm cut} = 0.1,0.2$). The sharp peak observed in the first bin of $\Delta R_{\rm STD-D^0}$ comes from single-track jets, where the charm hadron does not radiate at all. It is shown in Ref. 6 that the contribution of single-track jets is sensitive to the dead cone. The small $\Delta R_{\rm STD-D^0}$ region also contains jets with very few or very soft splittings off the charm quark, which do not significantly tilt the jet axis with respect to the $\rm D^0$ direction. When the Soft Drop condition is applied, these jets are more likely to be groomed away. Fewer jets satisfy Eq.~\ref{eq:SD} at small $\Delta R_{\rm SD-D^0}$ when $z_{\rm cut}=0.1$, and fewer still when the grooming condition is stricter ($z_{\rm cut}=0.2$). At large $\Delta R_{\rm axis}$, where a harder splitting at a wider-angle must exist to pull the jet axis away from the core of the jet, grooming has a minimal effect.

\section{The $\rm D^0$-tagged Groomed-Jet Momentum Fraction}
\label{sec:zg}

The final measurement highlighted is the groomed jet momentum fraction, 
\begin{equation}
z_{\rm g} = \frac{p_{\rm T, g}}{p_{\rm T, c}+p_{\rm T,g}},
\end{equation}
of the first splitting which passes the Soft Drop condition (Eq.~\ref{eq:SD}). Here, $p_{\rm T, c}$ and $p_{\rm T, g}$ are the transverse momentum carried by the radiating charm quark and the emitted gluon, respectively. This observable converges to the $1 \rightarrow 2$ splitting function in QCD and, by following the branch containing the $\rm D^0$ meson during reclustering, we are specifically probing $\rm c \rightarrow cg$ emissions \cite{zg}. 

\begin{figure}
\centerline{\includegraphics[width=0.425\linewidth]{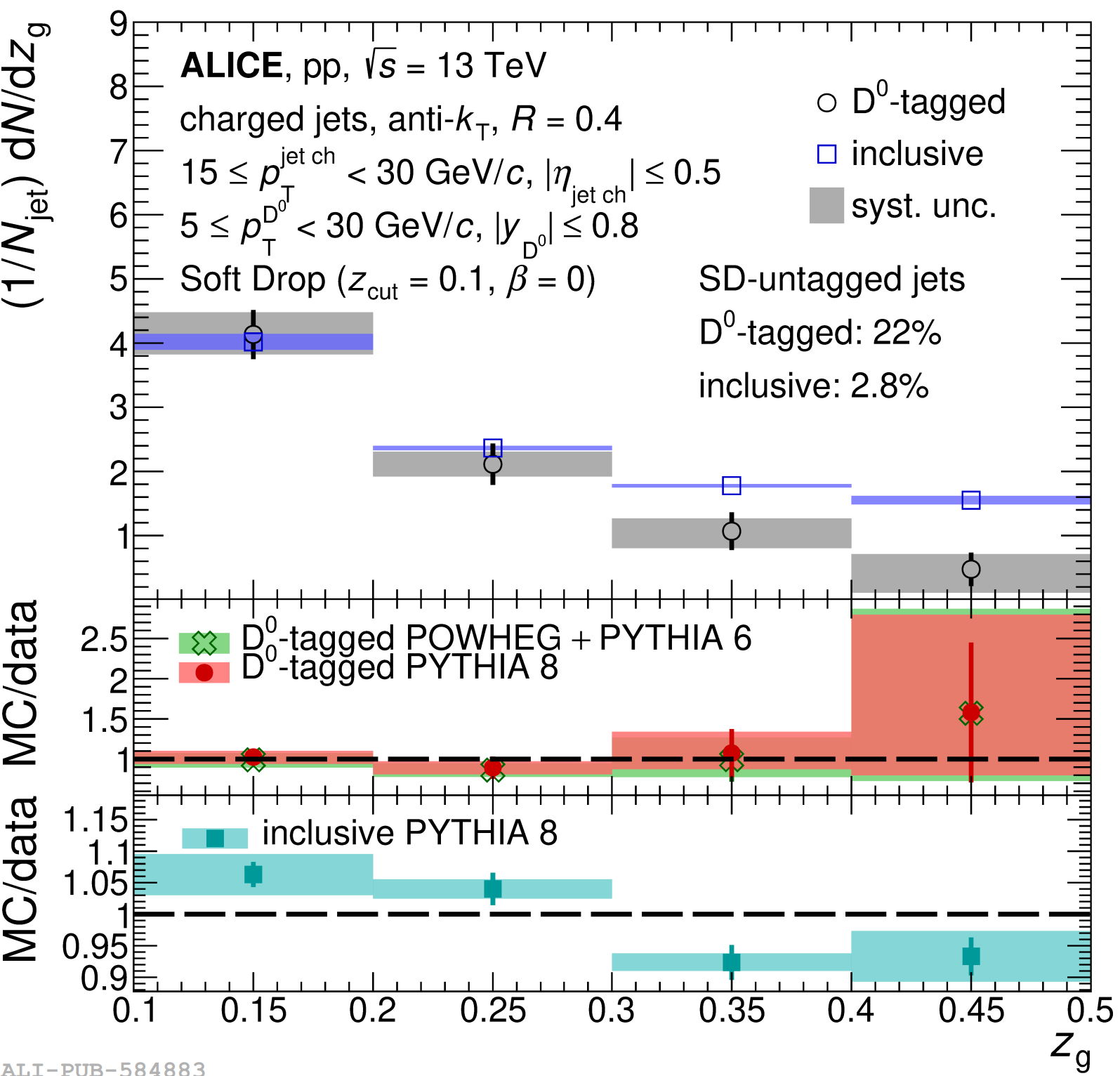} 
\includegraphics[width=0.425\linewidth]{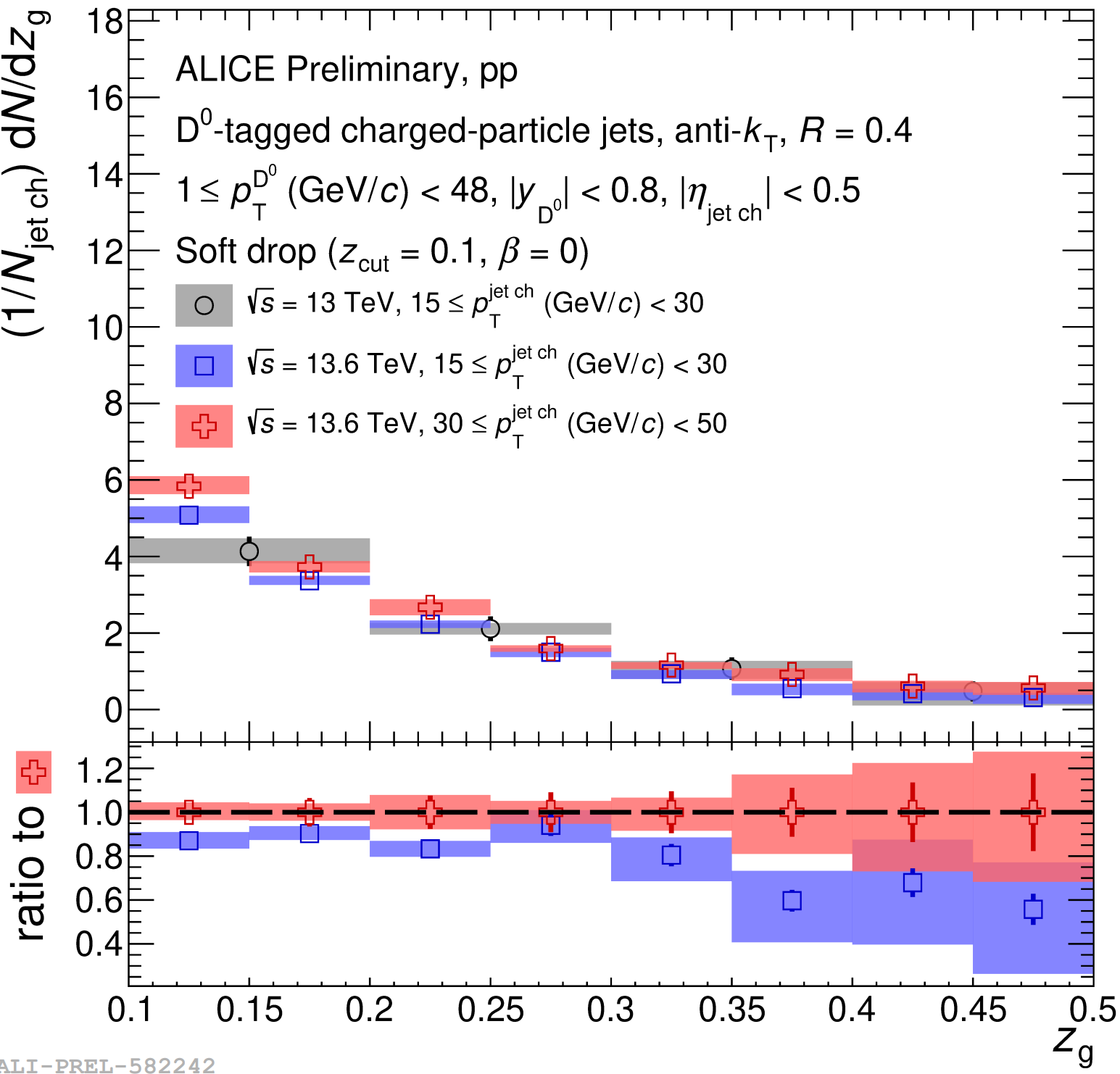}}
\caption[]{Fig.~\ref{fig:zg}(a): $\rm D^0$-tagged $z_{\rm g}$ (gray) compared to inclusive $z_{\rm g}$ (purple) in $15 < p_{\rm T}^{\rm ch~jet} < 30~{\rm GeV}/c$, calculated using $\sqrt{s}=13$ TeV Run 2 data. Fig.~\ref{fig:zg}(b): $\rm D^0$-tagged $z_{\rm g}$ from Fig.~\ref{fig:zg}(a) (gray), compared to new results from $\sqrt{s}=13.6$ TeV Run 3 data in $15 < p_{\rm T}^{\rm ch~jet} < 30~{\rm GeV}/c$ (blue) and $30 < p_{\rm T}^{\rm ch~jet} < 50~{\rm GeV}/c$ (red), which has an extended $p_{\rm T}^{\rm D^0}$ range and increased statistics \cite{zg}.} 
\label{fig:zg}
\end{figure}

In Fig.~\ref{fig:zg}(a), the $\rm D^0$-tagged distribution is steeper than the inclusive distribution, as heavy-flavor jets produce fewer symmetric (generally harder) splittings and are less likely to satisfy the Soft Drop condition. In Fig.~\ref{fig:zg}(b), the Run 3 data features improved precision and a broader kinematic range compared to Run 2. It also extends the measurement to a higher $p_{\rm T}^{\rm ch~jet}$ region, where the dead cone's inverse relation to the energy of the radiating quark leads to more jets passing the Soft Drop condition, increasing the total integral of the distribution.


\section{Conclusion}
ALICE has excellent heavy-flavor reconstruction capabilities at low jet energies, probing the region of phase space where the dead cone is most prominent. The suppression of small-angle radiation reduces parton splittings, lowering the EEC amplitude and increasing the contribution of single-track jets. Grooming removes jets with soft splittings, affecting more $\rm D^0$ jets than inclusive jets at low-$p_{\rm T}^{\rm ch~jet}$, and has less effect at high $p_{\rm T}^{\rm ch~jet}$ where the dead cone narrows. The $\rm D^0$ emerges as the leading particle at low-$p_{\rm T}^{\rm ch~jet}$, consistent with the steeper $\rm D^0$-tagged $z_{\rm g}$ distribution relative to the inclusive case. These findings motivate higher-precision measurements of charm baryons using data collected during the LHC Run 3 data-taking period, and similar studies in Pb--Pb collisions within this kinematic range.

\section*{References}

\begin{thebibliography}{99}
\bibitem{deadcone} ALICE Collaboration, \Journal{\it Nature}{7910}{605}{2022}.
\bibitem{PDG} S. Navas \textit{et al}. (Particle Data Group) \Journal{\PRD}{110}{030001}{2024}
\bibitem{anti-kT} M. Cacciari, G.P. Salam, G. Soyez, {\em JHEP} {\bf 04} (2008) 063.
\bibitem{eec} ALICE Collaboration, arXiv:2504.03431 [hep-ex], Apr 2025.
\bibitem{kyle_lee_EEC} E. Craft, K. Lee, B. Me\c{c}aj, I. Moult, arXiv:2210.09311 [hep-ph], Oct 2022.
\bibitem{jetaxes} ALICE Collaboration, arXiv:2504.02571 [hep-ex], Apr 2025.
\bibitem{SD} A.J. Larkoski, S. Marzani, G. Soyez, J. Thaler, {\em JHEP} {\bf 05} (2014) 146.
\bibitem{wta} A.J. Larkoski, D. Neill, J. Thaler, {\em JHEP} {\bf 04} (2014) 017.
\bibitem{zg} ALICE Collaboration, \Journal{\PRL}{19}{131}{2023}.
\end{thebibliography}


\end{document}